\documentclass[prl,showpacs,preprint]{revtex4}%
\usepackage{amsmath}
\usepackage{graphicx}
\usepackage{amsfonts}
\usepackage{amssymb}%
\begin{document}
\title{Numerical simulation of Bell inequality's violation using optical transverse
modes in multimode waveguides}
\author{Jian Fu, and Xun Zhang}
\affiliation{State Key Lab of Modern Optical Instrumentation, Centre for Optical and
Electromagnetic Research, Department of Optical Engineering, Zhejiang
University, Hangzhou 310027, China}

\begin{abstract}
We numerically demonstrate that \textquotedblleft mode-entangled
states\textquotedblright\ based on the transverse modes of classical optical
fields in multimode waveguides violate Bell's inequality. Numerically
simulating the correlation measurement scheme of Bell's inequality, we obtain
the normalized correlation functions of the intensity fluctuations for the two
entangled classical fields. By using the correlation functions, the maximum
violations of Bell's inequality are obtained. This implies that the two
classical fields in the mode-entangled states, although spatially separated,
present a nonlocal correlation.

\end{abstract}

\maketitle

Quantum entanglement is a fundamental concept and one of the most interesting
properties of quantum mechanics. The importance of quantum entanglement in
quantum information processing is by now widely appreciated
\cite{Nielsen,Bennett,Jozsa}. Quantum entanglement is first introduced by
Einstein, Podolsky, and Rosen (EPR) as most noticeable the EPR paradox
\cite{Einstein}, which is at the origin of quantum nonlocality. Bell proposed
a remarkable inequality imposed by a local hidden variable theory, which
enables an experimental test on the quantum nonlocality \cite{Bell}. The
violation of Bell's inequality indicates the presence of quantum entanglement.
Generally, the nonlocal properties, which appear in the correlation
measurement of quantum entangled states for two spatially separated particles,
are considered as the inherent features of quantum theory with no classical
analog. Numerous theoretical studies and experimental demonstrations have been
carried out to understand the nonlocal properties of quantum states. It is
noticeable that the violation of Bell's inequality is possible in nonlinear
optical processes such as multimode parametric amplifiers and four-wave
mixing, in which the output modes are in some way correlated \cite{Reid}. In
order to explore the intrinsic of quantum entanglement, some researchers yet
use classical optical fields to simulate quantum entanglement numerically and
experimentally \cite{Lee1,Hofer}.

The similarities between the Helmholtz equation and the Schr\"{o}dinger
equation have attracted some researches on the analogies between the
transverse modes in multimode waveguides and the quantum Fock states
\cite{Krivoshlykov,Nienhuis,Dragoman}. Besides the uncertainty relation
\cite{Marcuse}, Wigner phase-space distributions of the optical multimode
fields exhibiting negative regions are similar to quantum Fock states, which
had been verified experimentally \cite{Lee,Cheng}. Farther, some researchers
have argued convincingly that the classical Maxwell field plays the role of
the quantum wave function for a single photon \cite{Sipe}. This might be the
physical basis of the similarities. However, the similarities have been
limited principally to the measurement of first-order coherence, i.e.,
single-particle states. Classical wave analogs of the high-order nonlocal
coherence (quantum entanglement), i.e., multiparticle states, have been seldom
studied \cite{Lee1}. The researches on the high-order local coherence can date
back to the theoretical and experimental researches on interferences of two
independent laser beams in Refs. \cite{Magyar,Mandel}. Due to the random phase
difference between the two beams, no stable but transient interference fringes
can be observed. These local effects can be well explained by quantum theory
and classical electromagnetics. Recently, \textquotedblleft mode-entangled
states\textquotedblright\ based on the transverse modes of classical optical
fields propagating in multimode waveguides are proposed as the classical
simulation of quantum entangled states \cite{Fu}. The states can be regarded
as the nonlocal generalization of the high-order local coherence. It is
interesting that the mode-entangled states can also show the nonlocal
correlations such as the violation of Bell's inequality and the nonlocal
properties of optical pulses' group delays. However, the nonlocal correlations
are generally considered as the inherent features of quantum entanglement.

In this letter, we numerically demonstrate that the mode-entangled
states\ violate Bell's inequality by completely using classical
electromagnetics. Numerically simulating the correlation measurement scheme
proposed in Ref. \cite{Fu}, we obtain the normalized correlation functions of
the intensity fluctuations for two entangled classical fields. Then the
correlation functions are substituted into Bell's inequality and the maximum
violations of Bell's inequality are obtained. Despite all that, the
mode-entangled states and the quantum entangled states, from a physical
viewpoint, are different because the measurements of classical fields and
quantum states are different. The research on this simulation may be
important, for it not only helps to understand the nonlocal properties of
quantum entanglement from a new viewpoint, but also arouses interest in a full
optical quantum computation scheme based on the transverse modes of classical
fields \cite{Fu1,Man'ko}. And besides, it may be attractive to other research
fields. Such as researches on matter waves of Bose-Einstein condensates,
similar to the scheme, a quantum computation scheme based on the transverse
modes of the guided matter waves \cite{Erika} is also interesting.

In the correlation measurement scheme of Bell's inequality proposed in Ref.
\cite{Fu}, two independent monochromatic classical fields are prepared to
input a CNOT gate as the control and the target fields, respectively. When the
control field is given at mode superposition states $\left(  \left\vert
\operatorname*{TE}\nolimits_{0}\right\rangle \pm\left\vert \operatorname*{TE}%
\nolimits_{1}\right\rangle \right)  /\sqrt{2}$ and the target field is given
at the mode $\left\vert \operatorname*{TE}\nolimits_{0}\right\rangle $ or
$\left\vert \operatorname*{TE}\nolimits_{1}\right\rangle $, the output states
of the CNOT gate are so called \textquotedblleft mode-entangled
states\textquotedblright,
\begin{align}
\left\vert \Phi_{1}^{\pm}\right\rangle  &  =\frac{1}{\sqrt{2}}\left(
\left\vert \operatorname*{TE}\nolimits_{0}\right\rangle _{c}\left\vert
\operatorname*{TE}\nolimits_{0}\right\rangle _{t}\pm\left\vert
\operatorname*{TE}\nolimits_{1}\right\rangle _{c}\left\vert \operatorname*{TE}%
\nolimits_{1}\right\rangle _{t}\right)  ,\label{eq1}\\
\left\vert \Psi_{1}^{\pm}\right\rangle  &  =\frac{1}{\sqrt{2}}\left(
\left\vert \operatorname*{TE}\nolimits_{0}\right\rangle _{c}\left\vert
\operatorname*{TE}\nolimits_{1}\right\rangle _{t}\pm\left\vert
\operatorname*{TE}\nolimits_{1}\right\rangle _{c}\left\vert \operatorname*{TE}%
\nolimits_{0}\right\rangle _{t}\right)  ,\nonumber
\end{align}
where subscripts $c$ and $t$ represent the control and the target fields,
respectively. The states in each waveguide are mode superpositions, which are
obviously different from product states,
\begin{align}
\left\vert \Psi_{2}\right\rangle  &  =\left\vert \Psi_{c}\right\rangle
\otimes\left\vert \Psi_{t}\right\rangle \label{eq2}\\
&  =\frac{1}{2}\left(  \left\vert \operatorname*{TE}\nolimits_{0}\right\rangle
_{c}\pm\left\vert \operatorname*{TE}\nolimits_{1}\right\rangle _{c}\right)
\left(  \left\vert \operatorname*{TE}\nolimits_{0}\right\rangle _{t}%
\pm\left\vert \operatorname*{TE}\nolimits_{1}\right\rangle _{t}\right)
,\nonumber
\end{align}
where $\left\vert \Psi_{c}\right\rangle $ and $\left\vert \Psi_{t}%
\right\rangle $ represent the states of control and target fields,
respectively. The difference can be obtained not by measuring the single
field, but by the correlation measurement of the control and the target
fields. The correlation measurement can present that the two entangled fields
are inseparable to some extent, which is a nonlocal correlation.

To perform the correlation measurement of Bell's inequality, the control and
the target fields output from the CNOT gate are sent to spatially separated
(independent) mode analyzers (MAs), each of which contains a Y splitter and a
variable phase modulator $\theta_{1}$ ($\theta_{2}$). Based on the correlation
analysis, we get the correlation function,
\begin{equation}
S\left(  \theta_{1},\theta_{2}\right)  =\left\langle \hat{A}_{c}\left(
\theta_{1}\right)  \hat{B}_{t}\left(  \theta_{2}\right)  \right\rangle ,
\label{eq3}%
\end{equation}
where $\hat{A}_{c}\left(  \theta_{1}\right)  $ and $\hat{B}_{t}\left(
\theta_{2}\right)  $ are the intensity difference operators of the MAs'
outputs for the control and target fields, respectively,
\begin{align}
\hat{A}_{c}\left(  \theta_{1}\right)   &  =\hat{I}_{c}^{+}\left(  \theta
_{1}\right)  -\hat{I}_{c}^{-}\left(  \theta_{1}\right) \label{eq4}\\
&  =e^{2i\theta_{1}}\left\vert \operatorname*{TE}\nolimits_{0}\right\rangle
_{c}\left\langle \operatorname*{TE}\nolimits_{1}\right\vert _{c}%
+e^{-2i\theta_{1}}\left\vert \operatorname*{TE}\nolimits_{1}\right\rangle
_{c}\left\langle \operatorname*{TE}\nolimits_{0}\right\vert _{c},\nonumber\\
\hat{B}_{t}\left(  \theta_{2}\right)   &  =\hat{I}_{t}^{+}\left(  \theta
_{2}\right)  -\hat{I}_{t}^{-}\left(  \theta_{2}\right) \nonumber\\
&  =e^{2i\theta_{2}}\left\vert \operatorname*{TE}\nolimits_{0}\right\rangle
_{t}\left\langle \operatorname*{TE}\nolimits_{1}\right\vert _{t}%
+e^{-2i\theta_{2}}\left\vert \operatorname*{TE}\nolimits_{1}\right\rangle
_{t}\left\langle \operatorname*{TE}\nolimits_{0}\right\vert _{t},\nonumber
\end{align}
with $\hat{I}_{c}^{\pm}\left(  \theta_{1}\right)  $ and $\hat{I}_{t}^{\pm
}\left(  \theta_{2}\right)  $ being the MA's operations on the control and the
target fields, respectively. Apparently, there is no correlation between the
independent measurements of the two fields in the product states, hence, their
correlation function can be written as the product of the expected values,
\begin{align}
S_{\Psi_{2}}\left(  \theta_{1},\theta_{2}\right)   &  =\left\langle \Psi
_{2}\right\vert \hat{A}_{c}\left(  \theta_{1}\right)  \hat{B}_{t}\left(
\theta_{2}\right)  \left\vert \Psi_{2}\right\rangle \label{eq5}\\
&  =\left\langle \Psi_{c}\right\vert \hat{A}_{c}\left(  \theta_{1}\right)
\left\vert \Psi_{c}\right\rangle \cdot\left\langle \Psi_{t}\right\vert \hat
{B}_{t}\left(  \theta_{2}\right)  \left\vert \Psi_{t}\right\rangle .\nonumber
\end{align}
However, the correlation function of the mode-entangled states can not be
written in such form because the results of the independent measurements of
the two entangled fields are correlated. Here Bell's inequality is given as
the criterion to distinguish the mode-entangled state and the product state.

Consider an ensemble (large collection) of the independent and identical
systems that are labeled by $\lambda$ that satisfies normalization condition $%
%TCIMACRO{\dint \nolimits_{\Lambda}}%
%BeginExpansion
{\displaystyle\int\nolimits_{\Lambda}}
%EndExpansion
\rho\left(  \lambda\right)  d\lambda=1$, where $\rho\left(  \lambda\right)  $
is a distribution function and $\Lambda$ is spanned by $\lambda$. In this
model, the correlation function can be rewritten as a normalized form%
\begin{align}
S\left(  \theta_{1},\theta_{2}\right)   &  =\left\langle A_{c}\left(
\theta_{1}\right)  B_{t}\left(  \theta_{2}\right)  \right\rangle \label{eq6}\\
&  =\int_{\Lambda}A_{c}^{\prime}\left(  \theta_{1},\lambda\right)
B_{t}^{\prime}\left(  \theta_{2},\lambda\right)  \rho\left(  \lambda\right)
d\lambda,\nonumber
\end{align}
with
\begin{align}
A_{c}^{\prime}\left(  \theta_{1},\lambda\right)   &  =\frac{A_{c}\left(
\theta_{1},\lambda\right)  }{\left[  \int_{\Lambda}A_{c}^{2}\left(  \theta
_{1},\lambda\right)  \rho\left(  \lambda\right)  d\lambda\right]  ^{1/2}%
},\label{eq7}\\
B_{t}^{\prime}\left(  \theta_{2},\lambda\right)   &  =\frac{B_{t}\left(
\theta_{2},\lambda\right)  }{\left[  \int_{\Lambda}B_{t}^{2}\left(  \theta
_{2},\lambda\right)  \rho\left(  \lambda\right)  d\lambda\right]  ^{1/2}%
},\nonumber
\end{align}
where $A_{c}\left(  \theta_{1},\lambda\right)  $ and $B_{t}\left(  \theta
_{2},\lambda\right)  $ are the correlation measurement results corresponding
to each $\lambda$. Using the property of the product state's correlation
function shown in Eq. (\ref{eq5}), Schwarz inequality $\left\vert S\left(
\theta_{1},\theta_{2}\right)  \right\vert \leq1$ and Eq. (\ref{eq6}), we
obtain
\begin{align}
\left\vert S\left(  \theta_{1},\theta_{2}\right)  -S\left(  \theta_{1}%
,\theta_{2}^{\prime}\right)  \right\vert  &  \leq1\pm\int_{\Lambda}%
A_{c}^{\prime}\left(  \theta_{1}^{\prime},\lambda\right)  B_{t}^{\prime
}\left(  \theta_{2}^{\prime},\lambda\right)  \rho\left(  \lambda\right)
d\lambda\label{eq8}\\
&  +\left[  1\pm\int_{\Lambda}A_{c}^{\prime}\left(  \theta_{1}^{\prime
},\lambda\right)  B_{t}^{\prime}\left(  \theta_{2},\lambda\right)  \rho\left(
\lambda\right)  d\lambda\right] \nonumber\\
&  =2\pm\left[  S\left(  \theta_{1}^{\prime},\theta_{2}^{\prime}\right)
+S\left(  \theta_{1}^{\prime},\theta_{2}\right)  \right]  ,\nonumber
\end{align}
then Bell's inequality (CHSH inequality \cite{CHSH}) is given explicitly as
\begin{equation}
\left\vert B\right\vert =\left\vert S\left(  \theta_{1},\theta_{2}\right)
-S\left(  \theta_{1},\theta_{2}^{\prime}\right)  +S\left(  \theta_{1}^{\prime
},\theta_{2}^{\prime}\right)  +S\left(  \theta_{1}^{\prime},\theta_{2}\right)
\right\vert \leq2, \label{eq9}%
\end{equation}
which implies that the violation of Bell's inequality never occurs for the
product states. On the contrary, if the violation occurs, there is a nonlocal
correlation between the two fields of the mode-entangled states. In Ref.
\cite{Reid}, Bell's inequalities are generalized for the case of fields or
arbitrary intensity and the violation of Bell's inequalities is shown to be
possible in a regime showing strong violation of Schwarz inequality. But the
Schwarz inequality, related to the phenomena of photon antibunching and
squeezing, is different from $\left\vert S\left(  \theta_{1},\theta
_{2}\right)  \right\vert \leq1$ satisfied by the correlation function.

Let us now numerically simulate the scheme shown in Fig. 1 to obtain the
correlation functions of the control and the target fields. For such purpose,
we have to simulate the two independent optical fields propagating in
waveguides. In Refs. \cite{Magyar,Mandel}, the two independent light beams are
replaced by two monochromatic optical fields with random phases $\varphi_{1}$
and $\varphi_{2}$ uniformly distributed in $[0,2\pi]$. Therefore, the two
independent optical fields propagating in the waveguides can be regarded as an
ensemble labeled by the phase difference $\lambda=\varphi_{1}-\varphi_{2}$,
which also is a random variable uniformly distributed in $[0,2\pi]$. Thus, a
simulation of each $\lambda$ is equivalent to an independent experiment of the
ensemble, then the integrals of the simulations describe the behaviors of two
independent optical fields. Although it requires infinite numbers of the
simulations in theory, we can assign $\lambda$ with discrete values in
practice to realize the ergodicity of $\lambda$ by finite numbers of the simulations.

The finite differential beam propagation method (FD-BPM) employed in our
numerical simulation is widely used in the analysis of optical waveguides and
well proved effective and reliable by a lot of researchers \cite{Yevick}.
Utilizing FD-BPM, the evolution of waveguide modes are clearly demonstrated
using intensity distributions. And the waveguide output intensities can be
obtained by the integral of the intensity distributions. In the simulation of
the scheme in Fig. 1, the two MAs are separated far enough to avoid disturbing
each other, and the numerical simulation result is shown in Fig. 2. For a
certain $\lambda$, we obtain the relation between the MAs' output intensities
and $\theta_{1}$ ($\theta_{2}$) by changing the refractive indexes of the
phase modulators. We find that the output intensities of MAs also change with
the phase difference $\lambda$. The reason of the fact is that
$\operatorname*{TE}\nolimits_{0}$ modes of the control field and the target
field in one of the Mach-Zehnder interferometer's arms, due to the phase
difference $\lambda$, interfere with each other when they are passing through
the directional couplers of the CNOT gate. The interference causes the
intensity in Kerr-like medium varying with $\lambda$, therefore the phase
shift via the Kerr effect also varies. Finally, the MAs' output intensities
change with $\lambda$ due to the phase shift. The intensity differences
$A_{c}\left(  \theta_{1},\lambda\right)  $ and $B_{t}\left(  \theta
_{2},\lambda\right)  $ versus $\theta_{1}$ and $\theta_{2}$ are shown in Fig.
3. While $\lambda$ varies in $[0,2\pi]$, $A_{c}\left(  \theta_{1}%
,\lambda\right)  $ and $B_{t}\left(  \theta_{2},\lambda\right)  $ fluctuate
around their mean values, and the ranges of the fluctuations change with
$\theta_{1}$ and $\theta_{2}$, respectively.

First, we obtain the random sequences $A_{c}\left(  \theta_{1},\lambda\right)
$ and $B_{t}\left(  \theta_{2},\lambda\right)  $ subjected to a sufficiently
long random sequence of $\lambda$, then substitute them into Eq. (\ref{eq6}),
and obtain the correlation function $S\left(  \theta_{1},\theta_{2}\right)  $
via summation of $A_{c}\left(  \theta_{1},\lambda\right)  $ and $B_{t}\left(
\theta_{2},\lambda\right)  $ instead of the integral of $\lambda$. The
correlation functions $S\left(  \theta_{1},\theta_{2}\right)  $ of $\left\vert
\Phi_{1}^{\pm}\right\rangle $ and $\left\vert \Psi_{1}^{\pm}\right\rangle $
are shown in Fig. 4. After substituting $S\left(  \theta_{1},\theta
_{2}\right)  $ into Eq. (\ref{eq9}), we obtain the maximum violation of Bell's
inequality, as shown in Table 1, where the maximum values of $\left\vert
B\right\vert $ are the average results of many $\lambda$'s sequences.
Considerable attention should be paid to that $S\left(  \theta_{1},\theta
_{2}\right)  $ shown in Fig. 4 are quite different from the correlation
functions $\cos\left(  2\theta_{1}\pm2\theta_{2}\right)  $ or $\sin\left(
2\theta_{1}\pm2\theta_{2}\right)  $ of the quantum entangled states,
therefore, $\theta_{1}$, $\theta_{1}^{\prime}$, $\theta_{2}$ and $\theta
_{2}^{\prime}$ of the maximum violation of Bell's inequality in Table 1 may
not be $\pi/8$, $-\pi/8$, $0$, and $\pi/4$. During the simulation, we find
that the granularity of $\lambda$ hardly influences the simulation results, so
that the $\lambda$ is assigned with 64 discrete values in $[0,2\pi]$. Farther,
we obtain the relation between the maximum of $\left\vert B\right\vert $ and
the refractive index of the phase modulator $\theta_{3}$, as shown in Fig. 5.

In this letter, we have numerically demonstrated that the mode-entangled
states generated by two independent classical fields propagating through the
CNOT gate are different from the product states. The difference is presented
in the correlation measurement of the two fields. For the mode-entangled
states, Bell's inequality is violated. However, the violation never occurs for
the product states. This implies that two classical fields in the
mode-entangled states, although spatially separated, present a nonlocal
correlation, which means the mode-entangled states can be really regarded as
the classical simulation of quantum entangled states. The relevant experiments
are necessary because they can prove the feasibility of our scheme, moreover,
they might open new perspectives for optical quantum computation and communication.

This work was supported by the National Natural Science Foundation of China
under Grant No. 60407003.

\begin{description}
\item \pagebreak

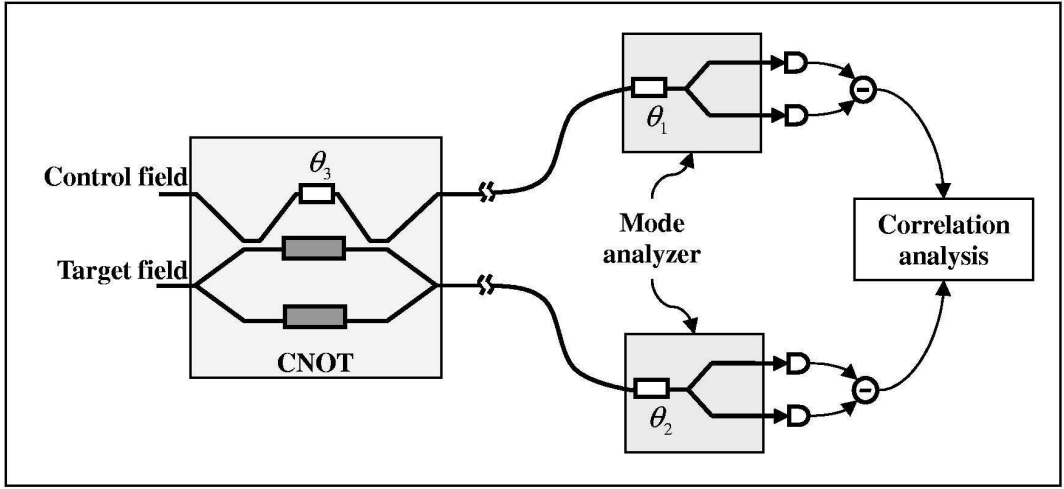
\item[Fig. 1:] Scheme of the numerical simulation.

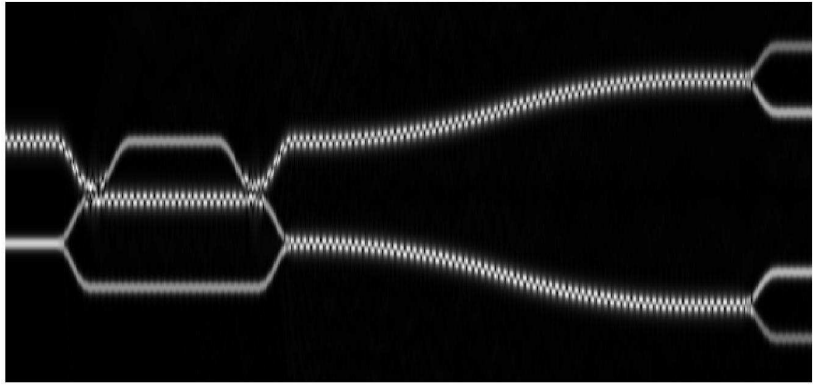
\item[Fig. 2:] BPM simulation result for the scheme shown in Fig. 1
corresponding to $\lambda=0$.

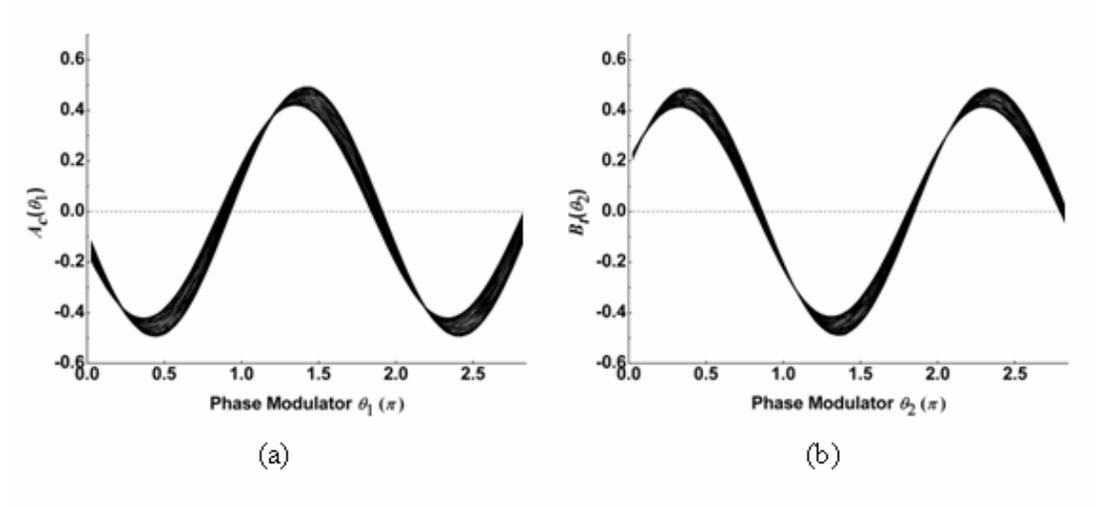
\item[Fig. 3:] The relation between the intensity differences and the phase
modulators: (a) $A_{c}\left(  \theta_{1},\lambda\right)  $ and $\theta_{1}$,
(b) $B_{t}\left(  \theta_{2},\lambda\right)  $ and $\theta_{2}$.

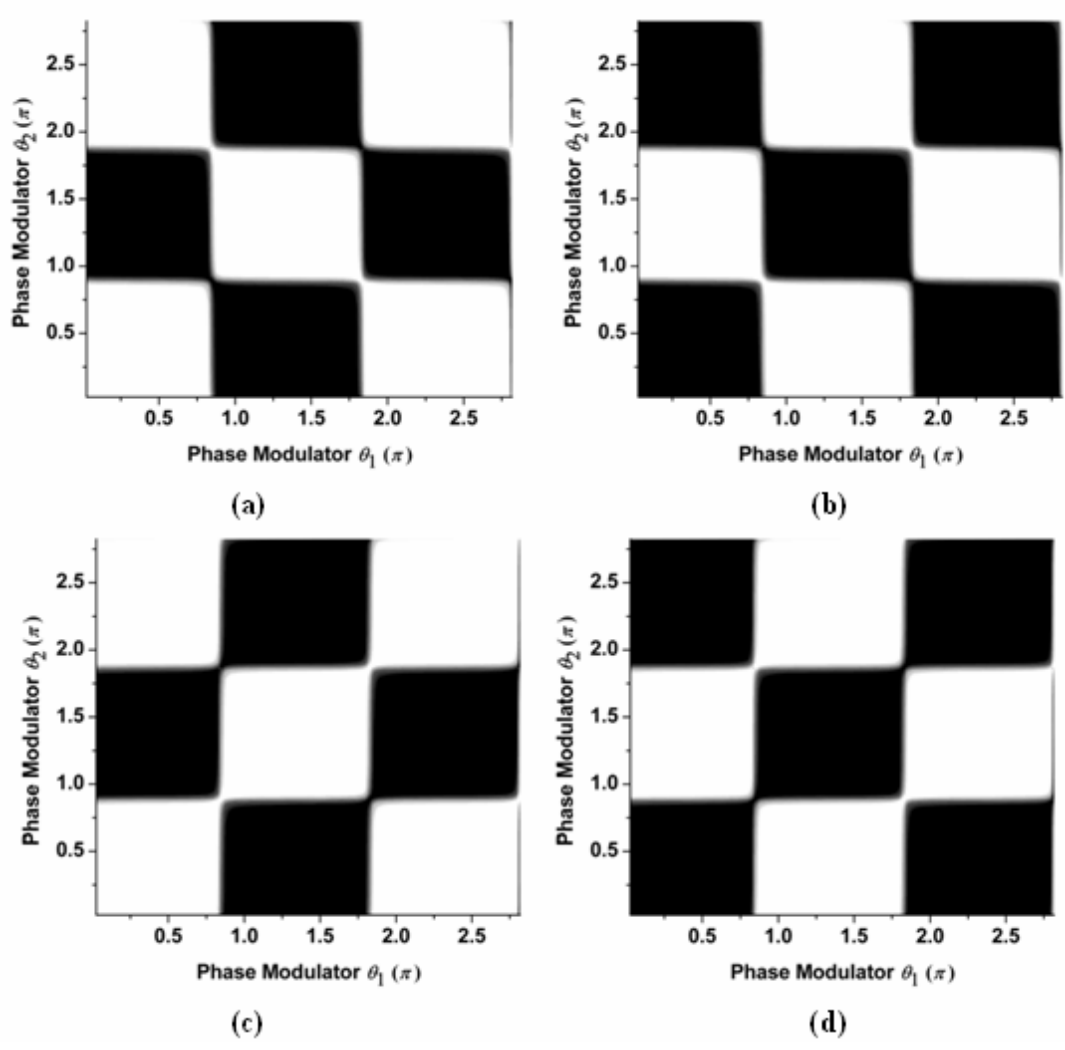
\item[Fig. 4:] The correlation functions $S\left(  \theta_{1},\theta
_{2}\right)  $ for mode-entangled states: (a) $\left\vert \Phi_{1}%
^{+}\right\rangle $, (b) $\left\vert \Phi_{1}^{-}\right\rangle $, (c)
$\left\vert \Psi_{1}^{+}\right\rangle $, and (d) $\left\vert \Psi_{1}%
^{-}\right\rangle $.

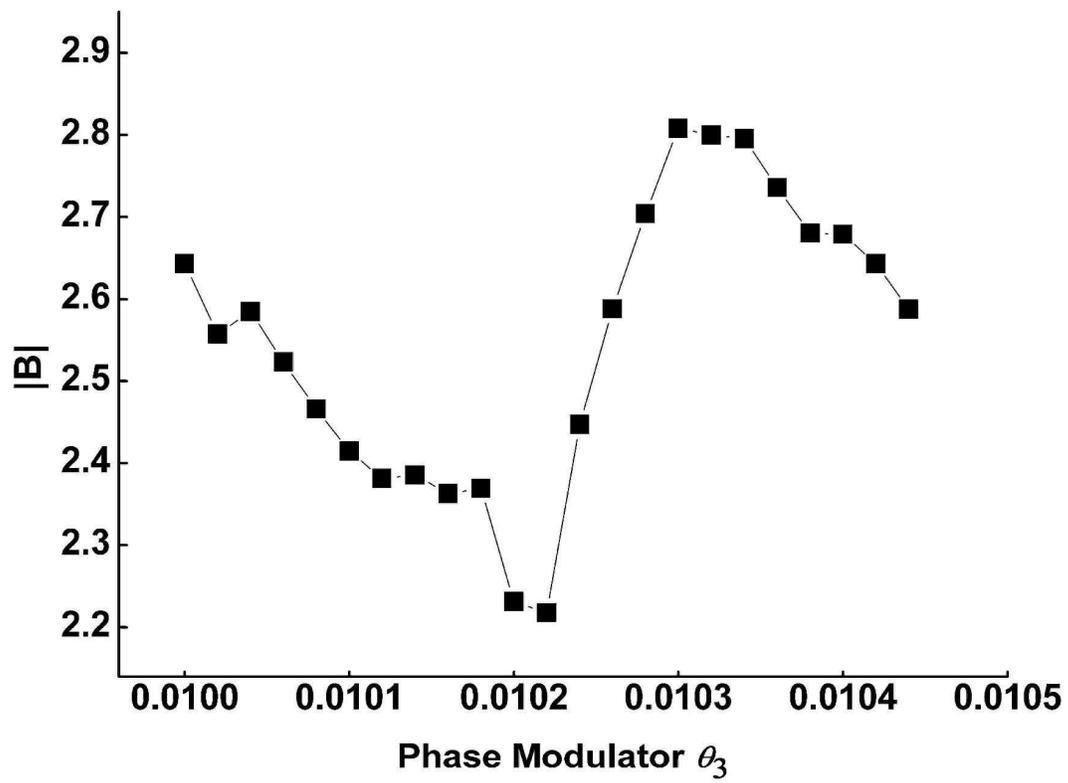
\item[Fig. 5:] The maximum of $\left\vert B\right\vert $ versus the refractive
index of the phase modulator $\theta_{3}$.

\item[Tabel 1:] 

\item
\begin{tabular}
[c]{|l|l|l|}\hline
$\left\vert \Phi_{1}^{+}\right\rangle $ & $\theta_{1}=\frac{77}{40}\pi
,\theta_{1}^{\prime}=\frac{35}{40}\pi,\theta_{2}=\frac{73}{40}\pi,\theta
_{2}^{\prime}=\frac{36}{40}\pi$ & max$\left\vert B\right\vert =2.7834$\\\hline
$\left\vert \Phi_{1}^{-}\right\rangle $ & $\theta_{1}=\frac{35}{40}\pi
,\theta_{1}^{\prime}=\frac{38}{40}\pi,\theta_{2}=\frac{111}{40}\pi,\theta
_{2}^{\prime}=\frac{113}{40}\pi$ & max$\left\vert B\right\vert =2.8041$%
\\\hline
$\left\vert \Psi_{1}^{+}\right\rangle $ & $\theta_{1}=\frac{83}{40}\pi
,\theta_{1}^{\prime}=\frac{75}{40}\pi,\theta_{2}=\frac{112}{40}\pi,\theta
_{2}^{\prime}=\frac{74}{40}\pi$ & max$\left\vert B\right\vert =2.8084$\\\hline
$\left\vert \Psi_{1}^{-}\right\rangle $ & $\theta_{1}=\frac{78}{40}\pi
,\theta_{1}^{\prime}=\frac{35}{40}\pi,\theta_{2}=\frac{72}{40}\pi,\theta
_{2}^{\prime}=\frac{113}{40}\pi$ & max$\left\vert B\right\vert =2.8086$%
\\\hline
\end{tabular}

\end{description}

\end{document}